\documentclass[aps,prl,superscriptaddress,reprint,nofootinbib,preprintnumbers]{revtex4-1}
\pdfoutput=1
\usepackage[normalem]{ulem}
\usepackage{epsfig}
\usepackage{dcolumn}
\usepackage{bm}
\usepackage{amssymb}
\usepackage{amsmath}
\usepackage{hyperref}
\usepackage[normalem]{ulem}
\usepackage{bm}
\usepackage{graphicx}
\usepackage[dvipsnames]{xcolor}
\usepackage{color}
\usepackage{float}

\usepackage{rotating}
\usepackage{bigints}
\usepackage{lineno}
\usepackage[normalem]{ulem}  % \sout{old text} for strikeout
\usepackage{ulem}
\usepackage{comment}
\usepackage{braket}

\usepackage{soul}

\usepackage{mathtools}

\def\th{\tau_{\text{hydro}}}

\def\r{\rho}
\def\w{w}

\begin{document}
\title{Critical slowing down and bulk viscosity in binary neutron star mergers}

\author{J.M. Karthein}
\email{jmkar@mit.edu}
\affiliation{Cyclotron Institute and Department of Physics and Astronomy,
Texas A$\&$M University, College Station, TX 77845, USA}
\affiliation{MIT Center for Theoretical Physics - a Leinweber Institute, Massachusetts Institute of Technology, Cambridge, MA 02139, USA}

\author{M. Pradeep}
\email{maneeshas1@iisc.ac.in}
\affiliation{Centre for High Energy Physics, Indian Institute of Science, Bangalore, 560012, India}

\author{R. Steinhorst}
\email{rstein99@mit.edu}
\affiliation{MIT Center for Theoretical Physics - a Leinweber Institute, Massachusetts Institute of Technology, Cambridge, MA 02139, USA}

\preprint{MIT-CTP/6014}

\date{\today}

\begin{abstract}
Hydrodynamic simulations of neutron star mergers rely on the clear separation between the strong-interaction, weak-interaction, and hydrodynamic timescales.
In this effective framework, weak Urca interactions are typically the slowest microscopic processes, and therefore the Urca rate determines the bulk-viscous dissipation. 
This assumed hierarchy of dissipative mechanisms can be decisively altered, without invalidating hydrodynamics, if the trajectory of the matter in a neutron star merger passes through the vicinity of a possible low temperature QCD critical point. The enhanced density fluctuations lead to critical slowing down and rapid growth of transport coefficients including bulk viscosity. While this growth is regulated by finite-time effects, finite-
size effects, and the breakdown of hydrodynamic scale separation, which bound the correlation length, we demonstrate that the QCD contribution to bulk viscosity can 
rival the electroweak contribution in realistic conditions. Thus,
critical dynamics could leave observable imprints on the hydrodynamic evolution of neutron star mergers. 
\end{abstract}


\maketitle

\section{Introduction}
\label{sec:intro}

Relativistic hydrodynamics describing a binary neutron star (BNS) merger is typically expected to have a hierarchy of timescales $\tau_{\rm strong} \ll \tau_{\rm weak} \ll \tau_{\rm hydro}$, suggesting that an ideal hydrodynamic description is sufficient. 
However, this hierarchy of scales is not guaranteed. For example, in the late inspiral and merger stage, we may instead have $\tau_{\rm hydro}\sim \tau_{\rm weak}$ \cite{Most:2022yhe}.
In such case, viscous effects can become important. 
A large bulk viscosity can have a non-trivial effect on gravitational wave emission~\cite{Most:2022yhe,Chabanov:2023blf}. Another thus-far-unexamined potential source of bulk viscous effects is critical slowing down (CSD) in the vicinity of a critical point (CP). Namely, if a CP exists in the phase-space trajectory of a BNS merger, 
a new timescale $\tau_{\rm CSD}$ emerges which corresponds to the relaxation time of the critical slow mode.
If $\tau_{\rm CSD}\sim \th$, critical effects may dominate viscous effects from electroweak physics or even lead to a breakdown of hydrodynamics, requiring an extended dynamical description \cite{Stephanov:2017ghc,Hohenberg:1977ym}.

There are several CP candidates relevant to this discussion. A well-established CP on the isospin-symmetric ($\mu_I=0$) QCD phase diagram with two light flavors of quarks is the nuclear liquid-gas phase transition that occurs around $\mu_B\approx\,900\, \text{MeV} , \, T\approx\,20\, \text{MeV}\,$ ~\cite{Fukushima:2013rx,Vovchenko:2016rkn}. Upon increasing the isospin asymmetry, the critical temperature is known to decrease, eventually dropping to zero for maximally isospin asymmetric matter \cite{Muller:1995ji, Ducoin:2005aa}. 
A higher-temperature critical point of the chiral symmetry restoration phase transition has long been discussed, but to date the evidence of its existence is not conclusive. 
Such a CP is conjectured to be
outside the phase space probed by BNS mergers,
with recent estimates from various methods suggesting $\mu_B\sim 600$~MeV \cite{Gao:2020fbl, Hippert:2023bel, Gunkel:2021oya, Basar:2023nkp, Clarke:2024ugt, Shah:2024img}, although the placement of this CP at a larger chemical potential and lower temperature than predicted is not ruled out. If this traditional CP exists, there is also the possibility of a second CP at a lower temperature marking another endpoint of the line of coexistence \cite{Hatsuda:2006ps,Yamamoto:2007ah,Abuki:2010jq}. This possibility remains true for all values of isospin asymmetry. 

Let us suppose a CP falls in the trajectory of the BNS merger system through the QCD phase diagram spanned by baryon ($\mu_B$) and isospin ($\mu_I$) chemical potentials and temperature ($T$). For each of the critical scenarios discussed above, the order parameter is a nearly conserved scalar dynamically coupled to the conserved momentum density. This candidate would belong to the static universality class of the 3D Ising model and the dynamic universality class  Model H~\cite{Son:2004iv,Hohenberg:1977ym}. In such case, whether critical fluctuations provide the dominant contribution to the hydrodynamic bulk viscous coefficient $\zeta$ (henceforth referred to as the bulk viscosity) will depend on the extent of critical slowing down as quantified by the correlation length $\xi$. While in equilibrium $\xi$ and consequently $\zeta$ is infinite at the CP, its growth is in practice limited by the finite duration allowed for equilibration and the finite system size.

The goal of this paper is to determine whether the presence of a QCD CP in the phase space probed by BNS mergers could significantly impact the merger bulk viscosity. We first discuss how finite time, finite size and the applicability of hydrodynamics constrain the growth of the correlation length near a CP. Next, we explain how this translates into a maximum critical enhancement of bulk viscosity, and use a characteristic choice for the non-critical part of the EoS to estimate this maximum critical contribution. Finally, we compare the bulk viscosity in the vicinity of the CP to typical contributions from electroweak process, and find that it is indeed conceivable that a critical contribution to bulk viscosity could far exceed the electroweak contribution, even in macroscopically large regions of the merger. 

\section{Bounds on the growth of the correlation length}
\label{sec:max-xi}

We will consider three factors that limit the growth of the correlation length. The first, and least relevant, is finite size; the spatial extent of a BNS merger system is of the order kilometers. The second is the finite time the system trajectory spends near the CP, in which time the correlation length cannot necessarily relax to its equilibrium value. 
The third is a breakdown of hydrodynamic scale separation when the correlation length becomes comparable to the length scale of temperature variation.

First, we consider finite time, following the example of Ref.~\cite{Stephanov:1999zu}. 
 Due to critical slowing down, the relaxation time for critical fluctuations goes as $\tau \sim \xi^z$. Consequently, we estimate that in finite time, $\tau$, $\xi$ has only relaxed to
\begin{align}\label{eq:finiteduration}
    \xi_{\rm max} \sim \xi_0 \left( \frac{\tau}{\tau_0} \right)^{1/z}\,.
\end{align}
where $\xi_0$ and $\tau_0$ are the correlation length and relaxation times far away from the CP. We will assume that $\xi_0\sim\tau_0\sim\Lambda_{\text{QCD}}^{-1}\sim 1$~fm.
Based on an approximate angular frequency of $\omega\sim 1$~kHz in the merger phase of BNS mergers, we 
estimate that a region of matter can spend about $\tau\sim 1$ ms near 
the CP during a BNS merger. In that case, we arrive at the estimate $\xi_{\rm max}\sim 10$~nm.

Now we shall estimate the bound on correlation length, $\xi$, obtained by the condition that critical enhancement of correlations cannot extend beyond the scale over which thermodynamic fields vary appreciably. Consider two spatial points, the local temperatures at these positions being $T_c$ and $T_c+\Delta T$ respectively. If these points are separated spatially so as to be identifiable as independent fluid cells in the coarse grained description, the coarse graining length scale $l$ is at most of the order of $\Delta T/ |\nabla T|$ where $|\nabla T|$ is the magnitude of spatial gradients of temperature in the vicinity of the CP. We will estimate for which value of $\Delta T$ the critical correlation length becomes comparable to $l$, i.e $\xi(\Delta T)\sim  \Delta T/ |\nabla T|$. This represents roughly the maximum achievable correlation length due to this constraint. 

To represent the universal variation of the correlation length in the vicinity of the CP, we perform a linear mapping between Ising variables (reduced temperature $r$ and magnetic field $h$) and QCD variables ($\mu_B, \mu_I, T$) as previously employed in the literature \cite{Parotto:2018pwx}. However, we choose the dimensionful scale $\mu_{B,c}$ rather than $T_c$, since this is the relevant scale in the low-temperature limit. We write
\begin{align} \label{eq:mapping}
    \frac{T-T_c}{\mu_{B,c}} &= \w (r \r \sin \alpha_1+ h \sin \alpha_2)\\
    \frac{\mu_B-\mu_{B,c}}{\mu_{B,c}} &= \w (-r \r \cos \alpha_1- h \cos \alpha_2)
\end{align}
where $\alpha_1$ and $\alpha_2$ are the angles the $r$ and $h$ axes make with the line $T=T_c$ at the CP, and $\rho$ and $w$ are two scale variables. These unknown mapping parameters are 
non-universal and change with $\mu_{I}$.  However, the Clausius-Clayperon relation requires that the slope of the phase boundary at $T=0$ is parallel to the $T$ axis. Therefore, for a small critical temperature, it is a reasonable choice to assume $\alpha_1\approx \pi/2$. For simplicity, we shall choose $\alpha_2=0$.\footnote{Our results make no distinction between $\alpha_1=-\pi/2,\alpha_2=0$ and $\alpha_1=\pi/2,\alpha_2=\pi$, meaning we describe equally well an upper or lower critical endpoint.}
With these choices, $T_c-T\propto r$ and $\mu_B-\mu_{B,c}\propto h$. 

In the 3D Ising universality class, the correlation length has a universal scaling $\xi/\xi_0=r^{-\nu}$ along the crossover line $h=0\,, r>0$ \cite{ZinnJustin}. Now, along a stretch of matter in the vicinity of the CP where $\mu_I$ and $\mu_B$ are approximately fixed, i.e along $h\approx 0$, we can see that $\xi_{\rm max}\sim l $ when
\begin{equation}
    \xi_0 \left( \frac{\Delta T}{\r\w \mu_{B,c}}\right)^{-\nu}\sim \frac{\Delta T}{|\nabla T|}\,.
\end{equation}
We can solve for $\Delta T\equiv T_c-T$ to get
\begin{equation} \label{eq:deltathydrobreakdown}
	\Delta T \sim \r\w \mu_{B,c} \left( \frac{\xi_0 |\nabla T|}{\r\w \mu_{B,c}} \right)^{1/(\nu+1)}\,,
\end{equation}
which suggests a maximum correlation length of 
\begin{align}
	\xi_{\rm max} &\sim l \sim \frac{\Delta T}{|\nabla T|}\\
	&\sim \xi_0  \left( \frac{\r\w \mu_{B,c}}{\xi_0 |\nabla T|}\right)^{\nu/(1+\nu)}\,.\label{eq:finiteextent}
\end{align}
Let us estimate that the temperature in the merger system varies on a scale of approximately $\frac{\partial T}{\partial x}=5$~MeV/km 
%\jmk{cite Most et al?}
(see e.g. Fig. 3 of \cite{Most:2022wgo}) and take $\r\w=1$ for the time being. Noting that in 3D Ising model universality class, we have $\nu/(1+\nu)\approx .386$, for a 
critical chemical potential of $\mu_{B,c}=1.2$~GeV, we estimate a maximum correlation length of $\xi_{\rm max}\sim 100$~nm, an order of magnitude greater than the finite time constraint (and therefore less stringent). 

These are, of course, mere order of magnitude estimates for a maximum conceivable value of $\xi$, but the possibility itself of a correlation length of $\mathcal{O}(10 \text{ nm})$ is striking, as it may suggest a modification of the electromagnetic or neutrino spectrum in the nanometer wavelength range. In this work, however, we will limit our investigation to critical enhancement of the bulk viscosity. 

\section{Critical Bulk Viscosity}\label{sec:zetacsd}

Here we will determine the bulk viscosity following the procedure of  
Ref.~\cite{Martinez:2019bsn}, making changes as necessary for our choice of mapping. 
The critical contribution to the bulk viscosity $\zeta_{\rm CSD}$ 
is thus obtained as 
\begin{equation}\label{eq:genericzeta}
    \zeta_{\rm CSD}(\omega) = c^2 \tau_0^2 \xi_0^2 T \int \frac{d^3k}{(2\pi)^3}\frac{2\chi_k^2}{-i\omega+2\Gamma_k}
\end{equation}
where we have $c=(e+P)\r\w \gamma\,$. 
Here $e$ is QCD energy density, $P$ is QCD pressure, $\gamma$ is the trilinear coupling between Ising energy density and the squared order parameter, $\omega$ is the driving frequency of density oscillations, $\Gamma_k$ is the relaxation rate of the order parameter, and $\chi_k$ is the susceptibility
\begin{equation}
    \chi_k = \frac{\xi_0^2(\xi/\xi_0)^{2-\eta}}{1+(k\xi)^{2-\eta}}\,.
\end{equation}
For convenience, and using $\gamma = \gamma_{\pm}^R r^{1-2\beta}$, we define $\tilde{c} = \r\w  \gamma_{\pm}^R$ to contain all information about the mapping apart from our choice of $\alpha_1,\alpha_2$.

Evaluating the integral in Eq.~\eqref{eq:genericzeta}, we arrive at an expression of the form
\begin{equation} \label{eq:zetaW}
    \zeta_{\rm CSD}(\omega) = \zeta_{\rm CSD}(0)\frac{I_K(\omega)}{I_K(0)} \,,
\end{equation}
where (in the Kawasaki approximation~\cite{Hohenberg:1977ym,Onuki:2002,Kawasaki:1970dpc})
\begin{equation}
    \label{eq:Ik}
    I_K(\omega) = \text{Re} \int_0^\infty dy \frac{y^2}{(1+y^{2-\eta})^2}\frac{1}{K(y)-i\omega/(2\Gamma_\xi)}.
\end{equation}
We will throughout assume a driving frequency of $\omega\sim 1\text{ kHz}$. Here $K(y)$ is the Kawasaki function (see \cite{Onuki:2002}), 
\begin{equation}
    K(y) = \frac{3}{4} \left[ 1+y^2+(y^3-1/y) \arctan(y) \right]\,,
\end{equation}
and $\Gamma_\xi$ is 
\begin{equation}\label{eq:GammaXi}
    \Gamma_\xi = \frac{T}{3\pi\eta_0 \xi^z}\,.
\end{equation}
Here and elsewhere in this work, $\eta_0$ refers to the non-critical shear viscosity.

The bulk viscosity at zero driving frequency in natural units is\footnote{This expression corrects a missing factor of $A^2\tau_0T$ from Eq.~(44) of \cite{Martinez:2019bsn}, which is otherwise the equivalent expression in the case $\alpha_1=0$.
}
\begin{align} \label{eq:zeta0}
    \zeta_{\rm CSD} (0) &= \frac{3}{\pi} 
    %(\gamma^R_\pm Aa_{n\epsilon})^2 
    \tilde{c}^2
    (\tau_0T)^2 \left(\frac{e+P}{T}\xi_0^3\right)^2\eta_0 I_K(0) \left(\frac{\xi}{\xi_0}\right)^{z-\alpha/\nu}
\end{align}
where it is straightforward to evaluate $I_K(0) \approx .649$. The enhancement of bulk viscosity in the vicinity of the CP is due to the divergence of the ratio $\xi/\xi_0$ in the limit of $r\rightarrow 0$ and $h\rightarrow 0$, which is limited by the bounds on the growth of correlation length discussed in the previous section. The frequency-dependent result can be written as 
\begin{align} \label{eq:zetatscaling}
    \zeta_{\rm CSD} (\omega) &= \left(\frac{\tilde{c}}{\pi} \right)^2 T^3
    (\tau_0^2\xi_0^3) \left(\frac{e+P}{T}\right)^2  \frac{I_K(\omega)}{\Gamma_\xi} \left(\frac{\xi}{\xi_0}\right)^{-\alpha/\nu}\,.
\end{align}
We can understand the frequency-dependence by noting that  $I_K(\omega)/\Gamma_\xi$ scales as 
\begin{align}\label{eq:ikscaling}
\frac{I_K(\omega)}{\Gamma_\xi}\sim
    \begin{cases} 
      \Gamma_\xi^{-1} & \Gamma_\xi \gg \omega \\[10pt] 
      \Gamma_\xi^{1/3} & \Gamma_\xi \ll \omega
   \end{cases}\,.
\end{align}
Furthermore, $I_K(\omega)/\Gamma_\xi$ has a resonant peak at $\Gamma_\xi \sim \omega$, i.e. $\zeta_{\rm CSD}$ as a function of $\xi$ has a maximum at finite $\xi$. 

The critical contribution to $(e+P)/T$ (appearing in Eqs.~\eqref{eq:zeta0} and~\eqref{eq:zetatscaling}) vanishes at the CP, and therefore near the CP, it is dominated by the non-critical part of the EoS.  
In the next section, we shall use a Hadron Resonance Gas (HRG) model to approximate this non-critical part of the EoS. 

\section{Excluded Volume HRG Model}
\label{sec:hrg}

Below $T_c$, strongly-interacting matter is hadronic; thus, the Hadron Resonance Gas is a low energy effective 
model
for QCD.
The ideal HRG model treats hadrons as non-interacting point particles. The excluded volume (EV) HRG is a commonly employed modification that takes into account the repulsive interactions between hadrons by delegating each hadron a volume $v$ \cite{Vovchenko:2016rkn}. We use this to model the non-critical part of the QCD EoS. The pressure ($p$) and other derivable thermodynamic quantities such as energy density ($e$), baryon number density ($n_B$) and entropy density ($s$) are calculated for this model. In general, the non-critical part of the EoS will vary with isospin chemical potential. However, for our calculations, we consider $\mu_I=0$ for simplicity. For our calculations, we include all hadrons from the recently developed hadronic list based on the Particle Data Group (PDG) database with improved agreement on thermodynamics, including the equation of state, from lattice QCD methods, known as PDG21+~\cite{SanMartin:2023zhv}. The shear viscosity for EV-HRG can also be calculated \cite{Noronha-Hostler:2012ycm,Gorenstein:2007mw,McLaughlin:2021dph}, and depends on the model parameter $r$, the effective radius of the hard sphere hadrons. We will use $r=0.1$~fm, which is a reasonably good match to lattice calculations of pressure and entropy at $\mu_B=0$ where they are available~\cite{McLaughlin:2021dph}. 

\begin{figure}
    \centering
    \includegraphics[width=\linewidth]{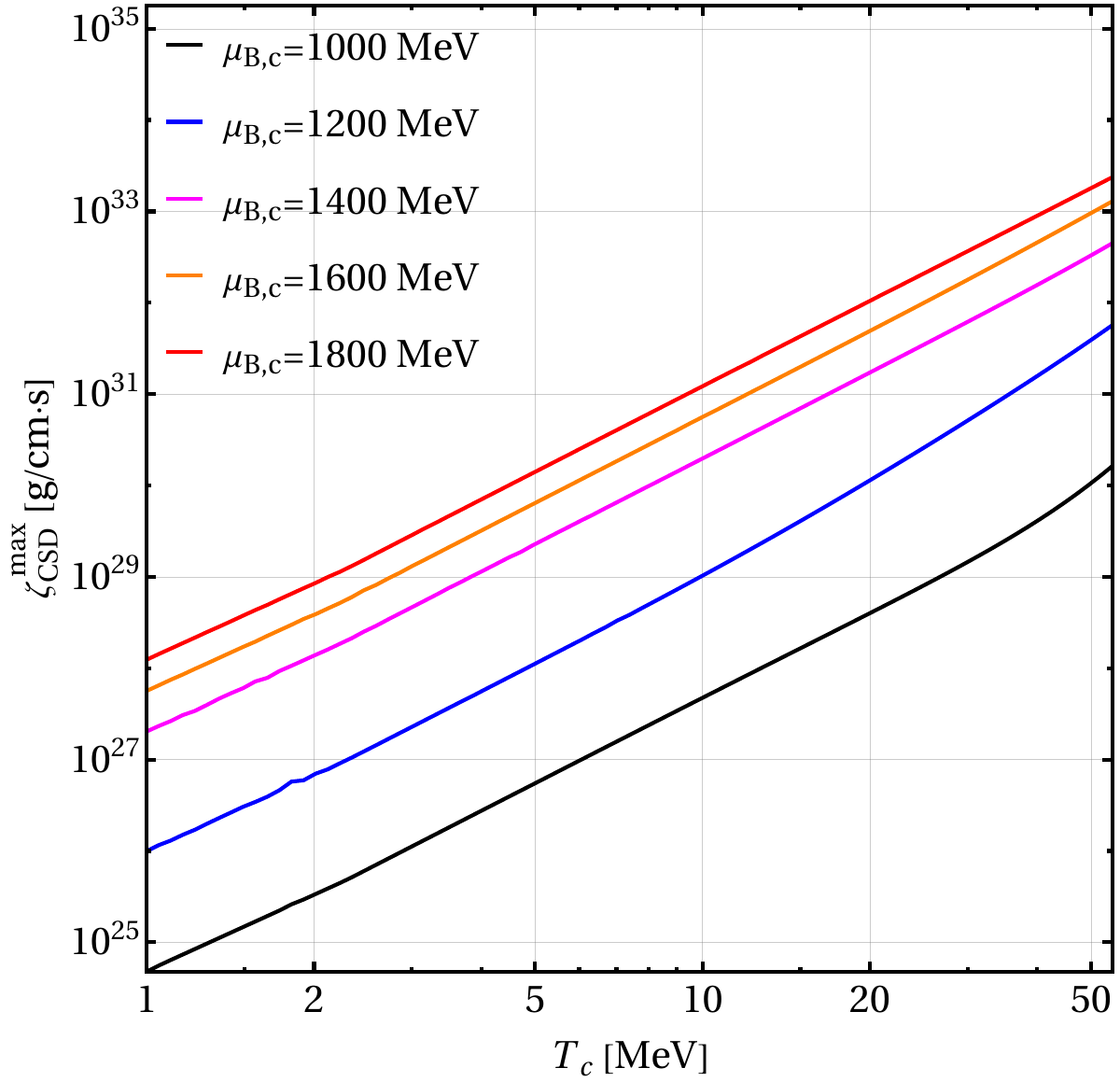}
    \caption{Estimated $\zeta_{\rm CSD}$ at the boundary of applicability of hydrodynamics
    for $\omega=1$~kHz, calculated using Eq.~\eqref{eq:zetatscaling} and an EV-HRG non-critical part of the EoS as described in the text. 
    The correlation length $\xi$ is taken to be the upper bound as suggested by the minimum of Eq.~\eqref{eq:finiteduration} and Eq.~\eqref{eq:finiteextent}, the two constraints discussed in the text. }
    \label{fig:zeta_scan}
\end{figure}

Now we can evaluate the bulk viscosity as shown in Eq.~\eqref{eq:zetatscaling} at the maximum conceivable correlation length $\xi_{\rm max}$ as determined by Eq.~\eqref{eq:finiteduration}. In this section, we will assume $\tilde{c}=1$ for simplicity, and defer analysis of the choice of $\rho\, w=\tilde{c}/\gamma^{+}$ to the next section. We perform a scan across temperatures and chemical potentials, shown in Fig.~\ref{fig:zeta_scan}. For the values of $T_c,\mu_{B,c}$ scanned, we estimate a maximum achievable bulk viscosity in the range $\zeta_{\rm CSD}^{\rm max}\sim 10^{25} - 10^{33}\text{ g/cm}\cdot\text{s}$. As we would expect from Eq.~\ref{eq:zetaW} and Eq.~\ref{eq:ikscaling}, we see a scaling $\zeta_{\rm CSD}^{\rm max}\sim T_c^{10/3}$ (since at low temperatures EV-HRG behaves as $e+P\sim T$).

Extensive work has been done calculating the bulk viscosity due to electroweak processes in various types of matter which may exist in neutron stars (see Ref.~\cite{Harris:2024evy} for a review). These works provide an electroweak baseline $\zeta_{\rm EW}$ to which we can compare the bulk viscosity due to QCD critical slowing down $\zeta_{\rm{CSD}}$. At temperatures above $\sim 10$~MeV, matter in the BNS merger becomes opaque to neutrinos~\cite{Alford:2018lhf}, and consequently $\zeta_{\rm EW}$ is lower. For a critical temperature above the neutrino-trapping temperature, 
we potentially have $\zeta_{\rm CSD}^{\rm max} \gg \zeta_{\rm EW}$ sufficiently close to the CP
(c.f. ~\cite{Alford:2022ufz}). However, it is important to note that $\zeta_{\rm CSD}^{\rm max}$ is only achievable very near the CP, and therefore may only be realized in a very small region of matter in the BNS merger. In the next section, we estimate the size of the region in which we might expect significant enhancement of the bulk viscosity, and find that it can be comparable to or larger than typical fluid cell sizes in BNS merger simulations.

\section{Relative Influence of Critical Enhancement}\label{sec:tempvariation}

In the previous sections, we demonstrated that it is possible to achieve enhancement of the correlation length $\xi$ by many orders of magnitude, and that this leads to a very large maximum achievable bulk viscosity. However, the interpretation of this enhanced viscosity in the context of a BNS merger simulation depends on the coarse-graining scale. If the correlation length becomes comparable to the coarse-graining scale, hydrodynamics breaks down, and $\zeta$ ceases to be well-defined. 
If the coarse-graining scale is too large, the region where $\zeta_{\rm CSD} \gtrsim \zeta_{\rm EW}$ is smaller than a single hydrodynamic cell. 
To properly understand the results of the previous section, we must examine the scaling of $\xi$ and $\zeta$ for temperatures near but not exactly at the CP.

First, we are interested in 
the temperature range over which
$\zeta_{\rm CSD} \gtrsim \zeta_{\rm EW}$.
As mentioned earlier, this range relies crucially on the value of the non-universal mapping parameter $\r\w$, which is unknown and depends on the microscopic details associated with the specific CP. 
In the analysis in the previous section, we used $\r\w=(\gamma_+^R)^{-1}$. Now we will allow $\r\w$ to vary as a proxy for our uncertainty about non-universal physics near the CP, and use $\gamma_+^R=0.43$~\cite{Martinez:2019bsn}, as we are studying thermodynamic ranges along the crossover line.

Next, we use universal scaling of the correlation length. The universal scaling $\xi=\xi_0 r^{-\nu}$ along $h=0$ can be re-expressed in terms of QCD variables using our chosen mapping as
\begin{equation}\label{eq:xitempscaling}
    \xi = \xi_0 (\r\w)^\nu \left( \frac{\Delta T}
    %T-T_c
    {\mu_{B,c}} \right)^{-\nu}\,.
\end{equation}
It follows from Eqs.~(\ref{eq:zetatscaling}-\ref{eq:xitempscaling}) that $\zeta_{\rm CSD}$ should scale as
\begin{equation}
    \zeta_{\rm CSD} \sim (\r\w)^2\left( \frac{\Delta T
    %T-T_c
    }{\r\w \mu_{B,c}}\right)^{-z\nu+\alpha}
\end{equation}
for $\Gamma_\xi \gg \omega$ as we move away from the critical temperature, where we note $-z\nu+\alpha \approx -1.8$. 

For each choice of $\r\w$ and $\mu_{B,c}$, we estimate a maximum correlation length $\xi_{\rm max}$ according to
\begin{equation}
\label{eq:ximax}
    \xi_{\rm max} = \min \left( \xi_0 \left( \frac{\tau}{\tau_0} \right)^{1/z}, \xi_0  \left( \frac{\r\w \mu_{B,c}}{\xi_0 |\nabla T|}\right)^{\nu/(1+\nu)}\right)\,,
\end{equation}
as in the previous section.
\footnote{ Here larger $\r\w$ means larger bulk viscosity; however, the dependence of other observables on $\r\w$ may be different, leading to a differing effective size of the critical region, see e.g.  \cite{Pradeep:2019ccv, Mroczek:2020rpm, Karthein:2025hvl}.} 
Then for a given $T$ and $\mu=\mu_{B,c}$, we can estimate $\xi$ from Eq.~\eqref{eq:xitempscaling}.  If $\xi<\xi_{\rm max}$, hydrodynamics remains applicable and we can consistently evaluate the bulk viscosity at the corresponding temperature and chemical potential.  If $\xi>\xi_{\rm max}$, hydrodynamics ceases to be valid in this region.

\begin{figure}
    \centering
    \includegraphics[width=\linewidth]{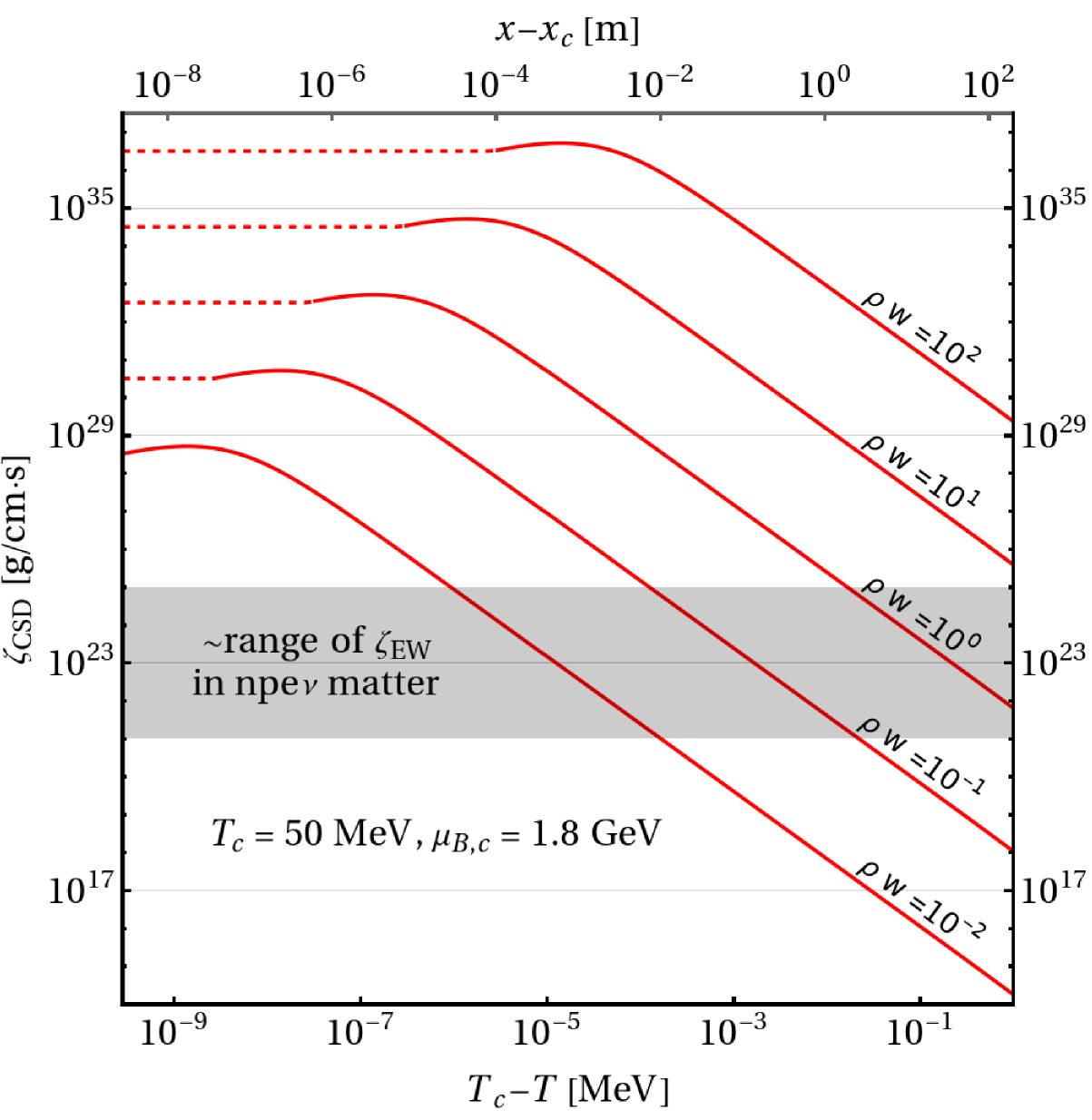}
    \caption{Bulk viscosity due to critical slowing down $\zeta_{\rm CSD}$ as a function of $T_c-T$ for $\omega=1$~kHz obtained from the critical scaling given by Eqs.~\eqref{eq:zetatscaling},~\eqref{eq:xitempscaling}. The gray band is based roughly on the neutrino-trapped calculations of Ref.~\cite{Alford:2021lpp}.
    We have chosen a CP at $\mu_{B,c}=\, 1.8 \, \text{GeV}, T_c= 50\,\text{MeV} , \, \alpha_{1}=\,\pi/2, \alpha_2=\, 0, \mu_{I}=0$, and use EV-HRG as a representative model of the on-critical part of the equation of state.
    Despite the power law dependence of $\zeta_{\rm CSD}$ on $\Delta T$, large enhancement of $\zeta$ can be present even at macroscopically relevant scales for $\r\w \gtrsim \mathcal{O}(1)$. The dashed horizontal segments of the curve are the regions of $T_c-T$ and $x-x_c$ where hydrodynamic description becomes invalid. As one can note, this happens only for coarse graining length scales which are orders of magnitude less than the typical fluid cell separation in neutron star merger simulations. 
    }
    \label{fig:dt}
\end{figure}

These results are plotted in Fig.~\ref{fig:dt}, where we fix $\mu_{B,c}=1.8 \text{ GeV}$ and $T_c=50\, \text{MeV}$. The lower horizontal axis shows the temperature deviation from the CP. The upper axis shows the subsequent spatial separation from the CP calculated assuming a uniform temperature gradient of 5 MeV/km (as we first did below Eq.~\ref{eq:finiteextent}). The gray region showing the approximate electroweak baseline $\zeta_{\rm EW}$ (assuming matter dominantly consisting of nucleons, electrons, and trapped neutrinos) 
makes clear that for this choice of $(T_c,\mu_{B,c})$, critical effects can be the dominant contribution for $\rho\w \gtrsim 1$. The dashed lines in the plot show the regimes where hydrodynamics breaks down. As evident from the figure, such a breakdown becomes relevant only if the coarse graining length scale is reduced below $10^{-4}\, \text{m}$ even for a relatively large value of $\r\w=10^{2}$, many orders smaller than typical coarse-graining length scales of 10s to 100s of meters in BNS mergers. From Eq.~\ref{eq:deltathydrobreakdown}, we find that hydrodynamics breaks down for a coarse graining length scale of 10 m only for $\r\w\gtrsim 10^6$.  
We note that a smaller choice of $T_c$ in the neutrino-transparent regime would result in a larger background viscosity $\zeta_{\rm EW}$; see Appendix A.

\section{Conclusions}\label{sec:concl}

We have investigated the impact of a potential QCD CP located near a hydrodynamic trajectory relevant for BNS merger simulations. We find that the growth of the correlation length $\xi$ in the proximity of this CP is limited primarily by the finite equilibration time within the merger system, and estimate that $\xi$ may grow to a maximum of $\sim \mathcal{O}(10)$~nm. 

Furthermore, we have performed an order of magnitude estimate for the critical enhancement of the bulk viscosity in matter near a CP within a BNS merger. Although this is dependent on the placement of the CP, the EoS, and the unknown non-universal combination $\r\w$, for a broad range of possible values, bulk viscosity arising from the critical physics can approach or exceed the electroweak contribution for $\r\w \gtrsim 10^0$.
This holds even at coarse-graining length scales of tens or hundreds of meters, which are of relevance for BNS merger simulations. Despite this significant critical effect, we also find that hydrodynamics remains a valid effective description at such coarse-graining scales, even when $\r\w$ is moderately large (i.e. $\rho w\lesssim 10^{6}$). 
This result is an interesting foray into the realm of a potential QCD CP search within compact astrophysical objects, a prospect which to our knowledge has not yet so far been seriously considered.

\section*{Acknowledgements}
We thank Mark Alford, Paulo Bedaque, Liam Brodie, Tom Cohen, Isabella Danhoni, Wojciech J. Jankowski, Krishna Rajagopal, Thomas Sch\"{a}fer, Enrico Speranza, and Mikhail Stephanov for useful discussions.  MP is supported via a Ramanujan Fellowship, Project File Number RJF/2025/000614. This work is supported by the U.S. Department of Energy, Office of Science, Office of Nuclear Physics under grant Contract Number DESC0011090. This research was supported in part by grant NSF PHY-2309135 to the Kavli Institute for Theoretical Physics (KITP)

\pagebreak

\appendix

\section{Appendix A: CP in the $\nu$-transparent regime}

\begin{figure}
    \centering
    \includegraphics[width=\linewidth]{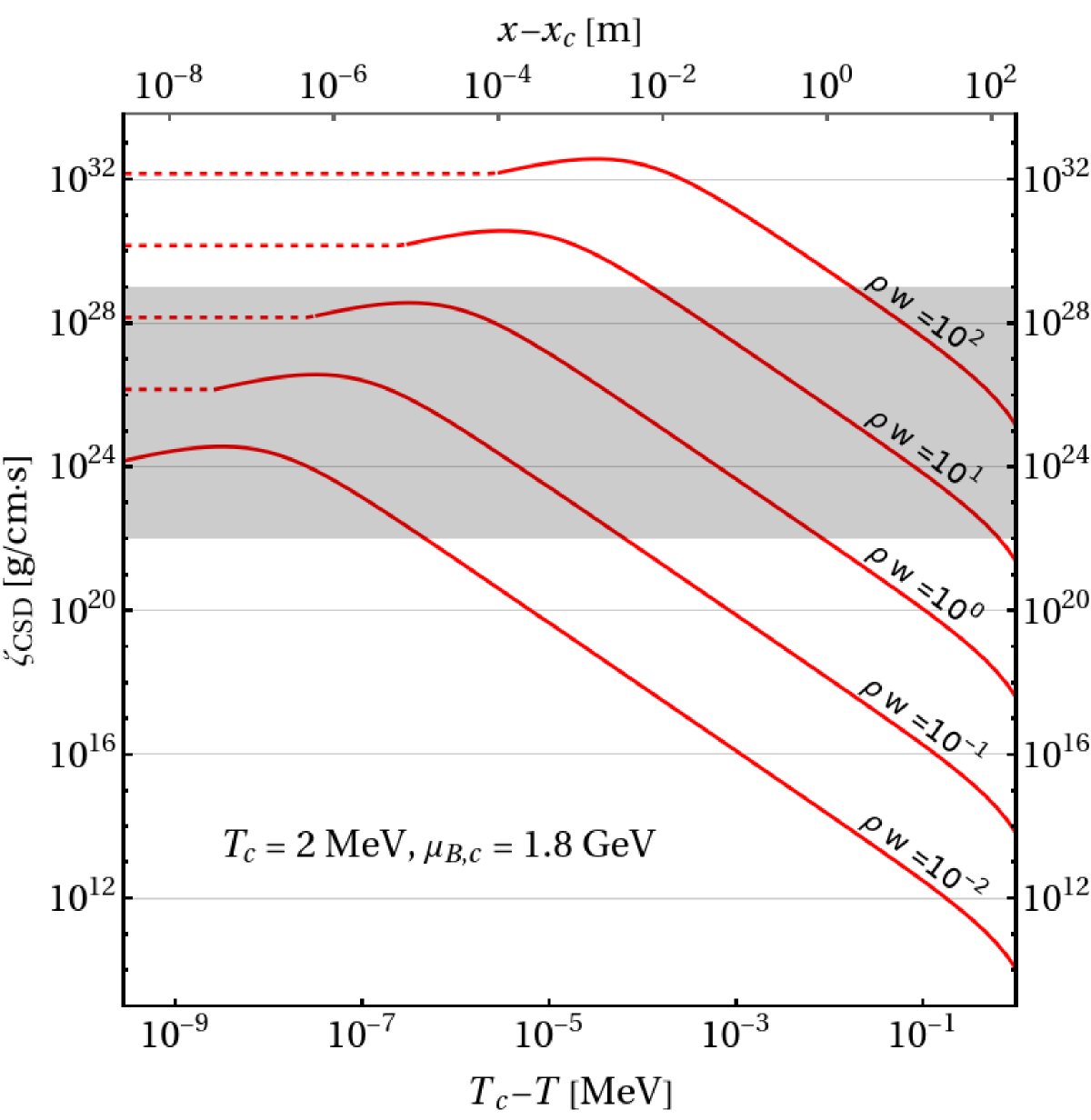}
    \caption{ As in Fig.~\ref{fig:dt}, $\zeta_{\rm CSD}$ as a function of $T-T_c$, but here with a critical temperature $T_c=2$~MeV. Here the gray band representing the approximate electroweak baseline is roughly based on the neutrino-transparent calculations in Ref.~\cite{Alford:2022ufz}. Comparing with Fig.~\ref{fig:dt}, one can see that a CP in the neutrino-transparent regime (e.g. $T_c=2$~MeV) has a much less distinct effect than one in the neutrino-trapped regime.
   } 
    \label{fig:smalltcdt}
\end{figure}

For completeness, we will repeat the analysis of Fig.~\ref{fig:dt} for a CP placed in the neutrino-transparent regime, $T_c=2\text{ MeV}$. Such a CP would be less prominent both because the critical contribution to $\zeta$ is smaller (see Fig.~\ref{fig:zeta_scan}) and because the smaller electroweak rate in the neutrino-transparent regime leads to an enhancement of the electroweak background \cite{Alford:2022ufz}. This can be seen in Fig.~\ref{fig:smalltcdt}. Nonetheless, for $\r\w \gtrsim 1$, there exists a region of the merger with meaningfully enhanced bulk viscosity. 

\bibliography{all}

\end{document}